\documentclass[aps,12pt,tightenlines,showpacs,amsmath,amssymb]{revtex4}
\newcommand{\be}{\begin{equation}}\newcommand{\ee}{\end{equation}}
\newcommand{\bea}{\begin{eqnarray}}\newcommand{\eea}{\end{eqnarray}}

\def\St{S^{\mathrm{tot}}}
\def\sys{\textsf{s}}
\def\DD{{\cal D}} \def\PP{{\cal P}} \def\WW{{\cal W}} \def\AAA{{\cal A}}
\def\KK{\overline K} \def\RR{\overline R} \def\subxy{_{xy}}
\def\SSnu{{\textsf{S}^\nu}} 

\begin{document}
\title{Generalized Clausius relation and power dissipation in non equilibrium stochastic systems}
\author{B. Gaveau}
\email{gaveau@ccr.jussieu.fr}
\affiliation{University Pierre et Marie Curie, Dept.\ of Mathematics, 75252 Paris Cedex 05, France}

\author{M. Moreau}
\email{moreau@lptl.jussieu.fr}
\affiliation{University Pierre et Marie Curie-CNRS, LPTMC (UMR 7600), case 121, 75252 Paris Cedex 05, France}

\author{L. S. Schulman}
\email{schulman@clarkson.edu}
\affiliation{Physics Department, Clarkson University, Potsdam, New York
13699-5820, USA}
\affiliation{Max Planck Institute for the Physics of Complex Systems, N\"othnitzer Str.\ 38, D - 01187 Dresden, Germany}

\date{\today}
\begin{abstract}
We extend certain basic and general concepts of thermodynamics to discrete Markov systems exchanging work and heat with reservoirs. In this framework we show that the celebrated Clausius inequality can be generalized and becomes an equality, significantly extending several recent results. We further show that achieving zero dissipation in a system implies that detailed balance obtains, and as a consequence there is zero power production. We obtain inequalities for power production under more general circumstances and show that near equilibrium obtaining maximum power production requires dissipation to be of the same order of magnitude.
\end{abstract}

\maketitle

Within the context of stochastic dynamics it turns out to be possible to make significant extensions of classical thermodynamics. For example, although the issue of power production is simply beyond thermodynamics (by definition), in the stochastic dynamical framework we are able to show that maximal power production entails a compulsory rate of dissipation (essentially entropy production) of the same order when close to equilibrium. In other results to be reported below, we generalize the Clausius relation and develop circumstances under which it is an equality.

We consider the stochastic dynamics of a discrete system $\sys$ undergoing a discrete time process. The elementary time step $\tau$ is taken as the time unit. The dynamics is defined by a stochastic matrix $R\equiv (R_{xy})$, where $R_{xy} \equiv  p(x,t+\tau \vert y,t)$ is the transition probability from $y$ to $x$ in time $\tau$. 

Let $\gamma=(y_{0},u_1, u_2, u_{N-1},x_0)$ be an $N$ step path from $y_0$ to $x_0$. The weight of this path is 
\be
\label{eq01}
W(\gamma)=R_{x_0 u_{N-1}} R_{u_{N-1} u_{N-2}} \dots R_{u_1 y_0}   \,.
\ee
The conditional average of a function $F(\gamma,t)$, given the initial and final states $y_0$ and $x_0$, is
\be
\label{eq02}
\langle F(\gamma,t=N\tau)\rangle_{x_0 y_0} 
     =\frac{\sum_{\gamma:y_0 \to x_0;|\gamma|=N}{W(\gamma)F(\gamma,t)}}
           {\sum_{\gamma:y_0 \to x_0;|\gamma|=N}{W(\gamma)}}
\ee
where $|\gamma|$ is the number of steps in the trajectory $\gamma$. For each elementary transition $y\to x$, such that $R_{yx}$ and $R_{xy}$ differ from 0, we suppose that the following relation holds 
\be
\label{eq03}
\frac{R_{xy}}{R_{yx}} =\exp (\delta \St)_{xy} 
\ee
where $(\delta\St)_{xy}$ is the total entropy variation of the system of interest, $\sys$, and of the other systems that are implied in the transition. Relation (\ref{eq03}), which can be deduced from microscopic detailed balance under appropriate conditions, has been previously used and discussed in the literature \cite{GS, GMS}.

We can write $(\delta\St)_{xy}=\delta_{xy} s +\delta S_{xy}$, where $s$ is the entropy of system $\sys$, $\delta_{xy}s$ is its variation from $y$ to $x$, and $\delta S_{xy}$ is the corresponding total variation of the entropy of other systems. For any function $h$ of the system state $x$, we write $\delta_{xy}h = h(x) - h(y)$. Thus $\delta_{xy}s = s(x) - s(y)$, whereas $(\delta S)_{xy}$ is the entropy variation of the external systems: it is supposed to depend only on $x$ and $y$, as discussed below, but in general it is \textit{not} the variation from $y$ to $x$ of any function of the system state alone \cite{GS, GMS}. If, under special conditions, it happens that there is a function $\St(x)$ such that $(\delta\St)_{xy} =  \St(x) - \St(y)$ for all $(x,y)$, the only stationary distribution $p^0(x)$ of $\sys$ is $p^{0}(x)\propto \St(x)$, and detailed balance \cite{carmichael, spohn, walls, evans, GMS} holds 
\be 
\label{eq04} R_{xy} \,p^0(x)\,=\,R_{yx} \,p^0(y) \,;
\ee
but this is not true in general. 

We assume that the system $\sys$ can exchange energy with several reservoirs $\SSnu$, labelled by indices $\nu= 1,2,\ldots$. Reservoir $\SSnu$ is characterized by its temperature $T_\nu$ or, better, by its inverse temperature $\beta_\nu=1/T_\nu$; the Boltzmann factor k$_{B}$ is always taken to be unity. The entropy variation of $\SSnu$ is - $\beta_\nu\delta q_\nu$ when it supplies energy (i.e., heat) $\delta q_\nu$ to the system. The system can also receive energy (i.e., work) $\delta w \equiv \delta q_0$ from a mechanical system $\textsf{S}^0$, whose entropy, by definition, does not change in time. We characterize the mechanical system by its zero inverse temperature, $\beta_0=0$.

During an elementary transition $y\to x$, we assume that the system can receive heat $\delta q_{xy}$ from \textit{at most one of the reservoirs}(whose inverse temperature is denoted $\beta _{xy})$, and work $\delta w_{xy}$ from the mechanical system. Thus the energy variation of the system is $\delta _{xy}e = \delta q_{xy}+\delta w_{xy}$. We suppose, as is currently done, that during an elementary transition $y\to x$ the mechanical work $\delta w_{xy}$ can be expressed as a function of $x$ and $y$ alone: then the same property holds for $\delta q_{xy}$ and for the entropy variation of the reservoir which interacts with $\sys$ during $y\to x$. The total entropy variation $\delta \St(\gamma ,t=N\tau)$ along an $N$-step trajectory $\gamma$ from $y_0$ to $x_0$ is
\begin{equation}
\label{eq10}
\delta \St(\gamma,t=\,N\tau)=
           \sum_{n=0}^{N-1} {\left[{\delta_{u_n u_{n+1}}s + \delta S_{u_n u_{n+1} }} \right]} 
           =\delta_{x_0 y_0} s 
               + \sum_{n=0}^{N-1} {\left[{-\beta_{u_{n+1}u_n} \delta q_{u_{n+1} u_n} } \right]} 
\end{equation}
Define $\bar \gamma$ to be the time reversal of the trajectory $\gamma$. Then using relation (3) it is straightforward to show that
\begin{equation}
\label{eq11}
\left\langle\exp(-\delta \St(\gamma,t=N\tau)\right\rangle_{x_0 y_0} 
=\frac{\sum_{\bar\gamma:x_0 \to y_0;\left|{\bar \gamma} \right|=N} {W(\bar \gamma)}}
      {\sum_{    \gamma:y_0 \to x_0;\left|      \gamma  \right|=N,}{W(     \gamma )}}
      =\frac{p(y_0,t|{x_0,0)}}{p(x_0,t|{y_0,0)}}
\end{equation}
Here $p(x,\tau|\,y,0)$ is the transition probability from $y$ to $x$ during time $t$. If $t/\tau = N \gg 1$, $p(x,\tau|\,y,0) \sim  p^0(x)$. Defining \cite{GMS, BMT}
the information potential of $x$ by $\phi (x)=-\log p^0(x)$ we obtain by (\ref{eq11})
\be
\label{eq18}
\left\langle \exp (-\delta \St(\gamma,t=N\tau)\right\rangle_{x_0 y_0} 
=\exp \left[{\phi(x^0)-\phi(y^0)} \right]
\equiv \exp\left[{\delta_{x_0y_0}\phi} \right]
\ee
Using the expression for $\delta \St$ we obtain our first main result, the \textit{generalized Clausius relation}
\be
\label{eq19}
\left\langle\exp \sum_{n=0}^{N-1} {\beta_{u_{n+1}u_n}} \delta q_{u_{n+1}u_n})\right\rangle_{x_0y_0}
=\exp[\delta_{x_0y_0} (s+\phi)]
\ee
which can also be written 
\begin{equation}
\label{eq20}
\left\langle
     \exp\left(\sum_{\beta_\nu>0} \beta_\nu \delta q^\nu\right)
         \right\rangle_{x_0y_0}
     =\exp[\delta_{x_0y_0} (s+\phi)]
\end{equation}
where $\delta q^\nu$ is the total heat received by the system from thermostat $\SSnu$ during the overall transition. From relation (\ref{eq20}) we obtain by Jensen's inequality
\begin{equation}
\label{eq21}
\left\langle\sum_{\beta_\nu>0} \beta_\nu\delta q^\nu \right\rangle_{x_0y_0} 
\le \delta_{x_0y_0}(s+\phi)
\end{equation}
Multiplying (\ref{eq21}) by the joint probability $p(y_{0},0;x_{0},t)$ and summing over $x_0$ and $y_0$ yields 
\begin{equation}
\label{eq22}
\left\langle\sum_{\beta_\nu>0} \beta_{\nu} \delta q^\nu \right\rangle
\le 
\Delta _t \overline s 
\end{equation}
where $\langle\cdot\rangle$ denotes the global average over all paths between times 0 and $t$, and $\Delta_t \bar s \equiv \bar s(t)-\bar s(0)$ is the variation of the macroscopic entropy $\overline s$ of the system between times 0 and $t$, with
\begin{equation}
\label{eq23}
\bar {s}(t)=
\left\langle s(x)-\ln [p(x,t)]\right\rangle
=\sum_x s(x) p(x,t)- \sum_x p(x,t)\ln [ p(x,t)]
\end{equation}

\textit{Inequalities} (\ref{eq21}) \textit{and} (\ref{eq22}) \textit{are changed into equalities if and only if in each elementary transition} $y\to x$ we have $\beta_{xy} \delta q_{xy} =\delta_{xy} (s+\phi)\equiv (s+\phi )(x)-(s+\phi)(y)$, which is easily shown to be equivalent to detailed balance, (4). Thus, detailed balance is the mesoscopic counterpart of the reversibility of a thermodynamic transition. 

Relation (\ref{eq22}) is the classical Clausius inequality \cite{clausius}, whereas (\ref{eq21}) is a mesoscopic version of this inequality. They can be compared with the generalizations of the Clausius inequality obtained in quite different contexts by Refs.\ \cite{shibata, muschik}. If we now restrict ourselves to an isothermal system, $\beta_\nu=\beta$ for any $\nu$, we easily recover the result of Ref.~\cite{GMS}
\begin{equation}
\label{eq30}
\left\langle \exp(-\beta\sum_{n=0}^{N-1} \delta w_{u_n u_{n+1}})\right\rangle_{x_0y_0 } 
=\exp [\delta _{x_0y_0} (-\beta f+\phi)]  \,,
\end{equation}
where $f(x) = e(x) - T s(x)$ is the mesoscopic free energy in state $x$. From relation (\ref{eq30}) one can recover the Jarzynski equality \cite{jarzynski}, which has given rise to a large literature in recent years (see for instance \cite{jarzynski, mauzerall, seifert, qian, crooks, hatano} and references therein).

Our previous results can be extended straightforwardly to inhomogeneous systems consisting of $n$ homogeneous cells, provided that during each elementary transition $x\to y$, each cell $k$ of the system $\sys$ interacts with at most one of the reservoirs, $\SSnu$, whereas it receives work from the mechanical system and energy from the other cells. The results can also be extended to a system exchanging various resources with the reservoirs: this more complex case will be addressed in a later publication.

As a particular consequence of the Clausius inequality, Carnot's theorem \cite{carnot} gives the maximum efficiency in work production for a motor operating between two heat baths only, which is attained if all transitions are reversible. Under these conditions, however, a finite transition needs an infinite time and power productions vanishes. In practice, \textit{power} is often the most relevant quantity and maximum efficiency may be less important. To address this problem, we now consider power and entropy production per unit time when the stochastic system $\sys$ is in its stationary state. The probability current corresponding to the elementary transition $y\to x$ is then
\begin{equation}
\label{eq32}
J_{xy} =R_{xy} p^0(y)-R_{yx}p^0(x)  \,.
\end{equation}
The stationary \textit{total entropy production} per unit time \cite{GS} can be written, thanks to~(3),
\begin{equation}
\label{eq33}
\DD=\frac12\sum_{x,y} {J_{xy}} \delta_{xy} \St
 = \frac12 \sum_{x,y} \left[R_{xy} p_0 (y) -R_{yx} p_0(x)\right]
\ln \frac{R_{xy} p_0(y)} {R_{yx} p_0(x)} \ge 0
\end{equation}
Relation (\ref{eq33}) expresses the well-known fact \cite{GS,BMT} that entropy production vanishes if and only if the stationary state satisfies detailed balance. Defining
\begin{equation}
\label{eq43}
\DD_{xy} \equiv J_{xy} \ln \frac{R_{xy} p_0 (y)}{R_{yx} p_0(x)} \ge 0  \,,
\end{equation}
we have
\begin{equation}
\label{eq44}
\DD=\frac12\sum_{x,y} \DD_{xy} 
   =\frac12\sum_{x,y} J_{xy} \delta_{xy} (s+\phi) 
            + \frac12 \sum_{x,y} J_{xy} \delta S^\nu_{xy} 
=\frac12\sum_{x,y,\nu \ne 0} J_{xy} \delta S^\nu_{xy}    \,,
\end{equation}
so that $\DD$ vanishes if $J_{xy} = 0$ for each transition during which $\sys$ actually interacts with one of the reservoirs. Since $\DD$ vanishes if and only if $J_{xy}=0$ for \textit{all} elementary transitions, we conclude that \textit{if} $J_{xy}= 0$ \textit{for each transition during which} $\sys$ \textit{actually interacts with a reservoir} $(\beta_{xy} \ne 0)$, \textit{then} $J_{xy}=0$ \textit{for all elementary transitions}. It is found from (\ref{eq03}) and (\ref{eq43}) that
\begin{equation}
\label{eq45}
\DD_{xy} =J_{xy} \left[\delta _{xy} (s+\phi) - \beta_{xy} \delta_{xy} e + \beta_{xy} \delta w_{xy} \right]     \,.
\end{equation}
From (\ref{eq45}) we deduce that the power received by $\sys$ is
\begin{equation}
\label{eq46}
\PP=\frac12\sum_{x,y} J_{xy} \delta w_{xy} 
   =\frac12\sum_{x,y,\beta_{xy}>0}
    {\left[{\frac{1}{\beta_{xy} }\DD_{xy} -\frac1{\beta_{xy} }J_{xy} \delta_{xy}(s+\phi )} \right]} 
\,.
\end{equation}
The first term on the right hand side is always non negative. It allows one to give a general, explicit definition of the \textit{power dissipation}
\be
\label{eq47}
\DD_\WW 
=\frac12\sum_{x,y,\beta_{xy}>0} \frac1{\beta_{xy}}\DD_{xy} 
=\frac12\sum_{x,y,\beta _{xy} >0}J_{xy} \frac1{\beta_{xy}}\ln \frac{R_{xy} p_0 (y)}{R_{yx} p_0(x)}
\ge 0   \,.
\ee
Thus the power $-\PP$ released by the system satisfies 
\begin{equation}
\label{eq48}
- \PP\le -\AAA \equiv \frac12\sum_{x,y,\beta_{xy}>0} {\frac1{\beta_{xy}} J_{xy}\delta_{xy}(s+\phi)} 
\,.
\end{equation}
This upper bound on $-\PP$ is obtained if and only if $\DD_{xy}= 0$ for any transition with $\beta_{xy}>0$, which implies that $J_{xy}=0$ for any transition: then detailed balance is satisfied and $\PP$ vanishes. Thus, in order that a system can act as a motor ($-\PP>0$), a necessary condition is that it is \textit{not in equilibrium}: \textit{the power dissipation should be positive}. 

Moreover, it is seen that $-\AAA$ is a linear function of the currents (if the stochastic potential is supposed to be fixed), whereas the power dissipation $\DD_{\WW}$ can be approximated by a quadratic function of the currents near detailed balance conditions. Under particular circumstances, these remarks allow us to make rough estimates of maximum power production and its relation to associated quantities. Because these results are quite suggestive we elaborate on the details. We suppose that the actual transition matrix, $R_{xy}$ is near in value to another, $\RR_{xy}$, which satisfies detailed balance. Moreover, both $R$ and $\bar R$ have the same stationary state, $p^0(x)$. Under these circumstances a remarkable fact emerges: an upper bound of the power delivered by the system is obtained if the dissipation is equal to the power produced. Let $K_{xy}\equiv R_{xy}p^0(y)$, with $\KK$ the corresponding quantity for $\RR$. Then by assumption $\KK_{xy}=\KK_{yx}$. To lowest order in the deviation of $R$ from $\RR$ one can easily show that
\be
-\DD_\WW\approx -\frac12\sum_{x,y,\beta_{xy}>0} \frac1{\beta_{xy}} \frac1{\KK_{xy}}\left(J_{xy}\right)^2
\,.
\label{eq50}
\ee
Writing $B_{xy}\equiv \delta_{xy}\left(s+\phi\right)$, we have
\bea
-\PP
&\approx&
\frac12 \sum_{x,y,\beta_{xy}>0} \frac1{\beta_{xy}}
      \left[
      -\frac1{\KK_{xy}}\left( J_{xy}-\frac12\KK_{xy}B_{xy}\right)^2
                    +\frac14\KK_{xy}\left(B_{xy} \right)^2
      \right]
      \nonumber\\
      &\le& \frac18 \sum_{x,y,\beta_{xy}>0}\frac{\KK_{xy}}{\beta_{xy}} \left(B_{xy} \right)^2
      =-\PP^{\mathrm{max}}
\,.
\label{eq51}
\eea
This upper bound is obtained if $J\subxy=\frac12\KK\subxy B\subxy$ for each transition, in which case the power dissipation is equal to the power produced:
\be
\DD^{\mathrm{max}}\approx 
     \frac18 \sum_{x,y,\beta_{xy}>0}\frac{\KK_{xy}}{\beta_{xy}} \left(B_{xy} \right)^2
      =-\PP^{\mathrm{max}}
\,.
\label{eq52}
\ee
In this situation, the power produced is half the quantity $-\AAA = -\PP + \DD$, given by Eqs.\ (\ref{eq46}) and (\ref{eq48}). It is clear that the currents must satisfy constraints which may not allow reaching this optimal situation. Nevertheless, maximizing $-\PP$ under the relevant constraints confirms that close to detailed balance, \textit{the maximum power released by the system is obtained when the power dissipation is of the same order of magnitude as the power produced.}

Of course, this may be invalid far from detailed balance conditions. More accurate, quantitative results should rely on specific examples, but in principle they can be obtained from the previous general tools. In this connection we mention a tantalizing example \cite{barger} from an elementary mechanics text. It is an exercise: ``Material drops from a hopper at a constant rate $dm/dt$ onto a conveyor belt moving with constant velocity $v$ parallel to the ground. What power motor would be needed to drive the belt?'' The answer turns out to be that the optimum power to be supplied is exactly twice the kinetic energy imparted to the particles (which can be seen by going into the belt reference frame). Thus the power output (the kinetic energy of the material) exactly equals the energy dissipated by friction. Our assumptions in the foregoing derivation are too restrictive to make this result a special case, but there is very much the suggestion that the factor $1/2$ that we have encountered is more general than our demonstration.

\end{document}